\documentclass[11pt]{article}

\input epsf
\usepackage{latexsym}
\usepackage{amssymb}
\usepackage[dvips]{graphicx}

\begin{document}

\centerline{\Large \bf Graviton production from extra dimensions}

\vskip 2cm

\centerline{K.~E.~Kunze\footnote{\tt Kerstin.Kunze@physik.uni-freiburg.de}}
\vskip 0.3cm
\centerline{\normalsize\sl
Fakult\"at f\"ur Physik, Universit\"at Freiburg,}
\centerline{\normalsize\sl
Hermann Herder Strasse 3, 
D-79104 Freiburg, Germany. }

\vskip 0.5cm

\centerline{{and}}

\vskip 0.5cm

\centerline{M.~Sakellariadou\footnote{\tt msakel@cc.uoa.gr}}
\vskip 0.3cm
\centerline{\normalsize\sl
Department of Astrophysics, Astronomy, and Mechanics,}
\centerline{\normalsize\sl
 Faculty of Physics, University
of Athens,}
\centerline{\normalsize\sl Panepistimiopolis, GR-15784 Zografos, Hellas.
}

\vskip 3cm

\centerline{\bf Abstract}

\vskip 0.5cm

\noindent
Graviton production due to collapsing extra dimensions is 
studied.
The momenta lying in the extra dimensions are taken into
account. 
A $D$-dimensional background is matched to 
an effectively four-dimensional standard radiation dominated
universe. Using observational constraints on the present 
gravitational wave spectrum, a bound on the maximal
temperature at the beginning of the radiation era is derived.
This expression depends on the number of extra dimensions, as well
as on the $D$-dimensional Planck mass. Furthermore, it is found  
that the extra dimensions have to be large.

\vskip 1cm

\noindent

\section{Introduction}
Our observable universe is four-dimensional. However fundamental
theories, like for example string theories, require for consistency
more than four space-time dimensions.  In the standard Kaluza-Klein
approach \cite{kk} the extra dimensions are compact and curled up to
very small size, roughly of the order of the Planck scale, ${\ell
}_P\sim 10^{-33} {\rm cm}$, implying that space-time is effectively
four-dimensional. However recently, with the discovery of brane
solutions to string/M-theory, the idea of large extra dimensions has
attracted much attention \cite{hw}.  In this last approach, it is
proposed that ordinary matter is confined to a three-dimensional
sub-manifold, a three-brane, which is embedded in a higher dimensional
space-time; the graviton on the other hand is allowed to propagate
freely through the whole space-time. Neglecting the brane tension,
i.e. the energy density per unit three-volume of the brane, and
considering compact dimensions, one re-introduces the Kaluza-Klein
picture.  However, in this case the extra dimensions do not have to be
small.  Newtonian gravity has been tested and found to hold down to
scales of order of 1 mm \cite{newgra}. Below this scale, gravity could
in principle be higher than four-dimensional. In higher dimensional
gravity, the four-dimensional Planck scale is no longer a fundamental
scale; the higher dimensional Planck scale, $M_{\rm D}$, becomes instead
the fundamental scale. This allows to explain the huge difference
between the electroweak and the four-dimensional Planck scale, known
as the hierarchy problem. Assuming for simplicity that the $n$ extra
dimensions form an $n$-torus which has the same radius $R$ in each
direction, and using Gauss' law, the $D$-dimensional and the
four-dimensional Planck masses, $M_4$ and $M_{\rm D}$ respectively,
are related by \cite{ADD}
\cite{ruba}
\begin{eqnarray}
M_4^2=R^nM_{\rm D}^{n+2}~,
\end{eqnarray}
where $R^n$ is the volume of the $n$ extra dimensions. Thus, taking
$M_{\rm D}$ to be of the order of the electroweak scale, $M_{\rm D}\sim
1 {\rm TeV}$, the huge difference between $M_4$ and the
electroweak scale can be explained as due to the large size of the
extra dimensions.

Here we are going to study the effects of time-varying extra
dimensions on the graviton production.  We assume that the background
space-time can be written as a direct product of an internal space,
i.e. the extra dimensions, and a four-dimensional external space-time.
Furthermore, during an initial phase, the internal space is
contracting and the four-dimensional external space-time is
expanding. At some time this is matched to a radiation dominated flat
Friedmann-Lema\^{\i}tre-Robertson-Walker (FLRW) universe with the internal
dimensions frozen at constant size.  Due to the changing background
metric, there will be particle production. Here we will be in
particular concerned with the production of gravitons. Using
observational bounds on graviton spectra it is possible to derive a
relation between the $D$-dimensional Planck mass and the temperature at
the time of transition.  Usually it is assumed that there are no
massive modes excited, i.e. momenta lying in the internal space are
not taken into account.  Here however, we are going to assume that
the internal momenta are excited and we are interested in their effect
on the spectral energy density in four dimensions.  In order to
determine this effect, the internal momenta are integrated out.  This
will lead to a final expression for the spectral energy density, which
will be just depending on the four-dimensional momenta.

The formalism to deal with metric perturbations has already been 
developed in Ref.~\cite{abe}.
Particle production in higher dimensional space-times has been
also the subject of Refs. \cite{GV} and \cite{giov}.

\section{Graviton production}
Consider a $D(=d+n+1)$-dimensional space-time with line element,
\begin{eqnarray}
{\rm d}s^2=a^2(\eta)\left[{\rm d}\eta^2-\delta_{ij}{\rm d}x^i{\rm d}x^j\right]
-b^2(\eta)\delta_{AB}{\rm d}y^{A}{\rm d}y^{B},
\label{metr}
\end{eqnarray}
where $d=3$, and the indices $i,j=1,..,3$ and $A, B=4,..,3+n$. Thus,
$a$ and $b$ denote the scale factors for the internal and external
spaces respectively, $\eta$ stands for the conformal time, and
$\delta_{ij}$ is the Kronecker symbol. Assume that the extra
dimensions are compact and, for simplicity, that they all have the
same size, i.e. $0\leq y^{A}\leq 2\pi r$.  Then the length-scale, $R$,
characterizing the physical size of the extra dimensions is $R=2\pi r
b$.

The higher dimensional phase is matched to a radiation dominated
universe. It is assumed that the extra dimensions are dynamical during
the first stage. At the onset of the radiation dominated era they
become static.

Hence, the evolution of the background will be parametrized as follows,
\begin{eqnarray}
a(\eta)&=&a_1\left(-\frac{\eta}{\eta_1}\right)^{\sigma}
\;\;\;\;\;\; , \;\;\;\;\;\;
b(\eta)=b_1\left(-\frac{\eta}{\eta_1}\right)^{\lambda}
\;\;\; ,\;\;\;
\mbox {for} \;\;\;\; \eta<-\eta_1\nonumber\\
a(\eta)&=&a_1\left(\frac{\eta+2\eta_1}{\eta_1}\right)
\;\;\;\; , \;\;\;\; 
b(\eta)=b_1
\;\;\;\;\;\;\;\;\;\; ,\;\;\;\;\;\;\;\;\;\;\;\;
\mbox {for} \;\;\;\; \eta\geq -\eta_1~.
\label{e2}
\end{eqnarray}
In the following we set $a_1=b_1=1$. In Figure 1 we present the 
evolution of the background for $n=8$ extra dimensions.
\begin{figure}[h]
\centerline{\epsfxsize=2.5in\epsfbox{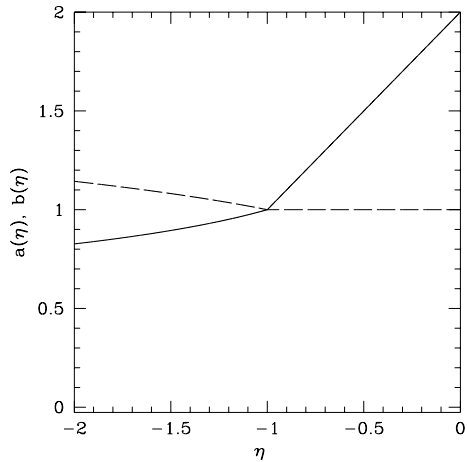}}
\caption{The evolution of the background during a
twelve-dimensional and an effectively four-dimensional
stage. Here the transition takes place at $\eta=-\eta_1=-1.$, and 
the number of extra dimensions is given by $n=8$. The solid line 
shows the scale factor $a(\eta)$ of the four-dimensional space-time
and the dashed line shows the scale factor $b(\eta)$ of the
eight-dimensional extra space.
}
\label{fig1}
\end{figure}

Within the context of the Kaluza-Klein theory, there are expected
oscillations of the $b$-field about the minimum. This is not
considered here, since this issue represents a question on its own. As
it can be seen from Eq.~(\ref{e2}), we consider a power-law behavior
for the $b$-field. Parametric resonance in the context of a higher
$D$-dimensional Kaluza-Klein theory has been only studied,
to our knowledge, for Kaluza-Klein modes ~\cite{param_res}.

The tensor perturbations of the $D$-dimensional
metric $G_{{\tilde A}{\tilde B}}$ take the form \cite{abe}
\begin{eqnarray}
\delta G_{{\tilde A}{\tilde B}}=
\left(
 \begin{array}{ccc}
 0 & 0 & 0\\
 0 & -a^2\gamma_{ij}(x,y) & 0\\
 0 & 0 & -b^2\gamma_{AB}(x,y)
 \end{array}
\right).
\nonumber
\end{eqnarray}
We consider that the unperturbed $D$-dimensional metric is a direct product,
thus there are no cross-terms in the above expression for the tensor
perturbations.
Only $\gamma_{ij}$ describes gravitational waves in the
four-dimensional space-time.
Although $\gamma_{AB}$ describes gravitational
waves in the $n$ extra dimensions, it is interpreted
as a scalar (matter) perturbation in the
four-dimensional space-time of the observer.
Both types of gauge-invariant tensor modes
satisfy the $D$-dimensional Klein-Gordon 
equation for a massless scalar field.
Here, we are only interested in the modes which behave as gravitational
waves in the four-dimensional space-time.
Writing $\gamma_{ij}=\phi(t,x,y)e_{ij}$, where
$e_{ij}$ is the polarization tensor,
the amplitude of the gravitational waves $\phi$ satisfies
the same equation as
a minimally coupled scalar field, i.e. \cite{abe} \cite{GW}
$$\Box\phi=0~,$$
 where $\Box$ denotes the d'Alembert operator in 
$D$-dimensions.  In the above background, in Fourier space, this yields to the
mode equation for the canonical field $\Phi=ab^{n/2}\phi$,
\begin{eqnarray}
\Phi_{l}^{''} + \left[ k^2 
+ \left(-{\eta\over\eta_1}\right)^{2\beta}q^2 
- {N\over \eta^2}\right] \Phi_{l}= 0~,
\label{e1}
\end{eqnarray}
where the three-vector ${\bf k}$ lies in 
the external space, the n-vector ${\bf q}$ is in the internal 
(compact) space, $l$ stands for $({\bf k}, {\bf q})$, and 
$$
\beta\equiv \sigma - \lambda
$$
$$
N\equiv \frac{(ab^{n/2})^{''}\eta^2}{ab^{n/2}}=
\sigma(\sigma-1)
      + n\sigma\lambda + n(n-2)\lambda^2/4
      + n\lambda(\lambda-1)/2~.
$$ 
Furthermore, the modulus of the external
and internal momentum vectors is denoted by $k\equiv\sqrt{|{\bf k}|}$ 
and $q\equiv\sqrt{|{\bf q}|}$, respectively.
Using the compactness of the extra dimensions the components 
$q^{A}$ can be written as $q^{A}={\cal N}_{A}/r$, where 
${\cal N}_{A}$ are integers.
This leads for the modulus of ${\bf q}$ to 
\begin{eqnarray}
q=\frac{{\cal N}}{r}~,
\label{q}
\end{eqnarray}
 with 
${\cal N}$ given by ${\cal N}=\sqrt{\sum {\cal N}_{A}^2}$.
For $\beta=-1$ exact solutions can be found, which are discussed in
Ref.~\cite{giov}. However, for general values of $\beta$ 
there is no known exact solution. Therefore, we follow 
Ref.~\cite{ax} to find an approximate solution.
In the following, it is assumed that $\sigma<0$ and $\lambda>0$, which
implies $\beta<0$. This describes a universe with a collapsing
internal (extra) dimensions and expanding external dimensions,
before the transition to an effectively four-dimensional space-time.

\subsection{Mode evolution in the presence of dynamical internal dimensions}
\noindent
During the higher dimensional state the background is
given by the vacuum Kasner metric. This constrains the
exponents $\lambda$ and $\sigma$, and thus $\beta$, and allows to write 
the metric as a function of the number of extra dimensions $n$.

Namely, the Kasner conditions $d\alpha_{\rm E}+n\alpha_{\rm I}=1$ and
$d\alpha_{\rm E}^2+n\alpha_{\rm I}^2=1$ imply for $d=3$ and $n\neq 0$, that
\begin{eqnarray}
\alpha_{\rm E}&=&\frac{1}{3}\left[
1-\frac{n}{3+n}\pm(-n)\sqrt{\frac{3}{n}\frac{2+n}{(3+n)^2}}
\right]\label{alp_E}\\
\alpha_{\rm I}&=&\frac{1}{3+n}\pm
\sqrt{\frac{3}{n}\frac{2+n}{(3+n)^2}}~.\label{alp_I}
\label{a_E}
\end{eqnarray}
These are related to the exponents $\sigma$ and $\lambda$ by
\begin{eqnarray}
\sigma&=&\frac{\alpha_{\rm E}}{1-\alpha_{\rm E}}\label{sig}\\
\lambda&=&\frac{\alpha_{\rm I}}{1-\alpha_{\rm E}}~.
\label{lambda}
\end{eqnarray}
Hence, calculating the parameter $N$ in Eq.~(\ref{e1})
yields $N=-(1/4)$, independently of the number of extra dimensions.
This can be understood by realizing that $ab^{n/2}=
(a^3b^n/a)^{1/2}=({\rm Volume}/a)^{1/2}$, where the volume is constrained
by the first Kasner condition.

The parameter $\beta$, which describes the difference in the expansion
rates of the internal and external space, takes values
$-1\leq\beta<-1/(1+\sqrt{3})$, where 
the lower limit corresponds to $n=1$ and the upper one 
to the limit of a large number of extra dimensions $n$.

The approximate solution of the mode equation, Eq.~(\ref{e1}),
is found by solving it in two different regimes.
Following the same approximation as in Ref.~\cite{ax}, two cases 
emerge.
\\

\noindent
{\bf Case (i):}
Assume that in Eq.~(\ref{e1}) the square of the
three-momentum $k$ always dominates over the term involving 
the internal momentum $q$. For very early times, $\eta\rightarrow-\infty$,
or equivalently very large $k$, hence very short wavelengths,
$k^2$ also dominates over the $\eta^{-2}$-term. 
And thus, the solutions are (in-coming) plane waves.
However, for times closer to the transition at $\eta=-\eta_1$, or 
for smaller $k$, hence longer wavelengths, 
$k^2<1/\eta^2$. Hence, this implies for the modulus of the
internal momentum $q$, that $q<(-\eta/\eta_1)^{-\beta}k$,
and thus together with $\eta_1\equiv 1/k_1$, the condition reads:
$q/k<(k/k_1)^{\beta}$.
Note that this case also includes the case $q=0$.
The in-coming state is effectively four-dimensional.
Then Eq.~(\ref{e1}) reduces to a Bessel equation which is
adequately, i.e. correctly normalized to an in-coming 
vacuum state, solved by
\begin{eqnarray}
\Phi_l=\sqrt{\frac{|k\eta|}{k}}H^{(2)}_{\mu}(|k\eta|)
\;\;\;\;\;\;\;\;\;\; \mbox {for}\;\;\;\;\;\;\;\;\;\;\;\eta<-\eta_1~,
\end{eqnarray} 
with $H^{(2)}_{\mu}$ the Hankel function of the second kind and
$\mu^2\equiv (1/4)+N$. Together with $N=-1/4$
the index of the Hankel function is identically vanishing,
$\mu=0$.
This solution is matched on super-horizon scales, i.e.
$k\eta_1\ll 1$,
to a radiation dominated universe at
$\eta=-\eta_1$. 
Thus, the canonical field $\Phi_l$ and its first 
derivative are given respectively by
\begin{eqnarray}
\Phi_{l}|_{\eta=-\eta_1}&\sim& -i\frac{2}{\pi}\sqrt{\eta_1}
\ln (k\eta_1)\nonumber\\
\Phi_l^{'}|_{\eta=-\eta_1}&\sim& -\frac{i}{\pi}\eta_1^{-\frac{1}{2}}
\left[2+\ln (k\eta_1)\right]~.
\label{phi_1}
\end{eqnarray}
\\

\noindent
{\bf Case (ii):}
Assume that the ${\bf q}$-term becomes dominant before
the perturbation in $D$ dimensions becomes super-horizon, hence
$k^2+q^2(-\eta/\eta_1)^{2\beta}>1/\eta^2$.
Thus, the mode equation can be approximated by
\begin{eqnarray}
\Phi_l^{''}+\left[k^2+q^2\left(-\frac{\eta}{\eta_1}\right)^{2\beta}
\right]\Phi_l=0,
\end{eqnarray}
which is approximately solved by (cf \cite{ax})
\begin{eqnarray}
\Phi_l\simeq\frac{\exp\left[i\eta
\sqrt{k^2+\kappa^2 q^2\left(-\frac{\eta}{\eta_1}\right)^{2\beta}}
\right]}
{\left(\frac{\pi}{2}\right)^{\frac{1}{2}}
\left[k^2+q^2\left(-\frac{\eta}{\eta_1}\right)^{2\beta}\right]^{\frac{1}{4}}},
\label{incom}
\end{eqnarray}
where $\kappa\equiv 1/(1+\beta)$, $\beta\neq -1$.
The case $\beta=-1$ can be solved exactly and it is 
discussed in Ref.~\cite{giov}.

At conformal time $\eta=\eta_q$ the internal momentum becomes dominant
over the external momentum and finally the $\eta^{-2}$-term dominates.
For $\eta>\eta_q$ the mode equation can be approximated by
\begin{eqnarray}
\Phi_l^{''}+\left[q^2\left(-\frac{\eta}{\eta_1}\right)^{2\beta}
-\frac{N}{\eta^2}\right]\Phi_l=0~,
\end{eqnarray}
which is solved by \cite{ax}
\begin{equation}
\Phi_l=c_q^{(1)}\sqrt{|q\eta|}H_{\mu\kappa}^{(1)}\left(|q\eta|
\kappa\left(-\frac{\eta}{\eta_1}\right)^{\beta}\right)
-i
c_q^{(2)}\sqrt{|q\eta|}H_{\mu\kappa}^{(2)}\left(|q\eta|
\kappa\left(-\frac{\eta}{\eta_1}\right)^{\beta}\right)~.
\label{intdom}
\end{equation}
Matching Eq.~(\ref{intdom}) to Eq.~(\ref{incom}) leads to
$c_q^{(1)}=0$ and $c_q^{(2)}=i/\sqrt{q}.$ 
Hence, for $|q\eta_1|\ll1$, i.e. super-horizon perturbations,
and using $H_0^{(2)}(z)\sim -i(2/\pi)\ln z$, one obtains that
$\Phi_l$ and its first derivative at $\eta=-\eta_1$ are given
respectively by
\begin{eqnarray}
\Phi_l\mid_{\eta=-\eta_1}&\sim& -i\frac{2}{\pi}\eta_1^{\frac{1}{2}}
\ln(\kappa q\eta_1)\nonumber\\
\Phi_l^{'}\mid_{\eta=-\eta_1}&\sim&
-\frac{i}{\pi}\eta_1^{-\frac{1}{2}}\ln(\kappa q\eta_1)
-i\frac{2}{\pi}(\beta+1)\eta_1^{-\frac{1}{2}}~.
\label{phii}
\end{eqnarray}

\subsection{Mode evolution with static internal dimensions}
The higher dimensional state evolves instantaneously into 
an effectively four-dimensional radiation dominated universe
at $\eta=-\eta_1$. The extra dimensions stay at a constant size 
after the transition, with scale factors given by Eq.~(\ref{e2}) for 
$\eta>-\eta_1$.
Then the mode equation, Eq.~(\ref{e1}), takes the form
\begin{eqnarray}
\Phi_l^{''}+\left[k^2+\left(\frac{\eta+2\eta_1}{\eta_1}\right)^2
q^2\right]\Phi_l=0~.
\end{eqnarray}
This can be transformed into the equation for
parabolic cylinder functions by
introducing $z\equiv(2q/\eta_1)^{1/2}(\eta+2\eta_1)$ and 
$a\equiv-(\eta_1/2q)k^2$ \cite{giov}.
This yields
\begin{eqnarray}
\frac{d^2\Phi_l}{dz^2}+\left(\frac{z^2}{4}-a\right)\Phi_l=0.
\label{diffz}
\end{eqnarray}
Its complex standard solution is given by \cite{AS}
\begin{eqnarray}
\Phi_l(z)={\cal B}\left[c_{-}E(a,z)
+c_{+}E^{*}(a,z)\right]~,
\end{eqnarray}
where for later convenience the constants
are chosen to be the Bogoliubov coefficients
$c_{\pm}$ and ${\cal B}$ is a normalization factor.
With the Wronskian condition on the mode functions
(see for example Ref.~\cite{GV}), i.e.
$\Phi_{l}^{*'}\Phi_{l}-\Phi_l'\Phi_l^{*}=i$,
and the normalization of the Bogoliubov coefficients
$|c_{+}|^2-|c_{-}|^2=1$, together with the
expression for the Wronskian of $E(a,z)$, $E^{*}(a,z)$
\cite{AS}, the normalization factor ${\cal B}$
is found to be 
${\cal B}=[\eta_1/(2q)]^{1/4}/\sqrt{2} $.


Hence the mode solution is given by
\begin{eqnarray}
\Phi(z)=\frac{1}{\sqrt{2}}\left(\frac{\eta_1}{2q}
\right)^{\frac{1}{4}}\left[c_{-}E(a,z)+c_{+}E^{*}(a,z)\right].
\end{eqnarray}
In order to make progress the solutions will be considered
in two regimes of approximation which will comprise 
cases (iii) and (iv).
\\

\noindent
{\bf Case (iii):} 
Assume that $|a|>z^2/4\;\;\;
\Leftrightarrow\;\;\;
k^2>[(\eta+2\eta_1)/\eta_1]^2 q^2$.
Using the limiting behaviour given in 
Ref.~\cite{AS} ([19.22.4]), $E$ can be approximated in this limit as
\begin{eqnarray}
E(a,\eta)\sim\sqrt{2}\frac{1}{2^{3/4}}\left|\frac{\Gamma(\frac{1}{4}
+\frac{1}{2}ia)}{\Gamma(\frac{3}{4}+\frac{1}{2}ia)}
\right|^{\frac{1}{2}}\exp\left[
-\frac{1}{4}\left(\frac{q}{k}\right)^2
\left(\frac{\eta+2\eta_1}{\eta_1}\right)^2\right]
e^{ik(\eta+2\eta_1)}
\end{eqnarray}
and a similar expression for the complex conjugate $E^{*}$.
Using
$$\left|\frac{\Gamma(\frac{1}{4}
+\frac{1}{2}ia)}{\Gamma(\frac{3}{4}+\frac{1}{2}ia)}
\right|^{\frac{1}{2}}\sim\left|\frac{a}{2}\right|^{-\frac{1}{4}}~,$$
the mode functions in this limit evolve according to
\begin{eqnarray}
\Phi_{l}\sim\frac{1}{\sqrt{2k}}\exp\left[
-\frac{1}{4}\left(\frac{q}{k}\right)^2
\left(\frac{\eta+2\eta_1}{\eta_1}\right)^2\right]
\left[c_{-}e^{ik(\eta+2\eta_1)}
+c_{+}e^{-ik(\eta+2\eta_1)}
\right].
\label{iii}
\end{eqnarray}
\\

\noindent
{\bf Case (iv):} 
Assume that $|a|<z^2/4\;\;\;\Leftrightarrow\;\;\;
k^2<[(\eta+2\eta_1)/\eta_1]^2/q^2$.
With the approximation of
$E(a,z)$ \cite{AS} in this limit the evolution of the modes is given by
\begin{eqnarray}
\Phi_l(\eta)\sim\sqrt{\frac{\eta_1}{2q(\eta+2\eta_1)}}
\left[c_{-}\exp\left(i\frac{q}{2\eta_1}(\eta+2\eta_1)^2\right)
+c_{+}\exp\left(-i\frac{q}{2\eta_1}(\eta+2\eta_1)^2\right)
\right]~.
\label{iv}
\end{eqnarray}

\subsection{Graviton spectra}
Gravitons are produced due to the changing background space-time.
Their spectral energy density might allow conclusions about the
early history of the universe. The Bogoliubov coefficients transform the
in-state into the out-state and determine the number of 
created particles. As it turns out there are 
three possible combinations of the four mode solutions 
discussed previously.
It is found that the evolution of modes  from the regime of validity 
of cases (i) and (iii), (i) and (iv), as well as (ii) and (iv) is
possible.
For these three regimes, the Bogoliubov coefficients have to be 
calculated. Then, the spectral energy density is
found by simply adding up the contributions from the
three parts. In order to find the total spectral energy density
in the effectively four-dimensional space-time after the 
transition, the momenta lying in the internal (extra) space are
integrated out. In this way the contribution of energy due to
the modes in the extra space can be estimated.

The Bogoliubov coefficients are calculated by matching the 
mode functions and their first derivatives at the transition time 
$\eta=-\eta_1$, on super-horizon scales.
The resulting expressions together with the appropriate 
range of validity are given below. 

\subsubsection*{Case (i) matched to (iii):}
\begin{eqnarray}
e^{ik\eta_1}c_{-}\sim-\frac{1}{\sqrt{2}\pi}
\frac{1}{\sqrt{k\eta_1}}
e^{\frac{1}{4}\left(\frac{q}{k}\right)^2}
\left[2+\left(1+\left(\frac{q}{k}\right)^2
+2ik\eta_1\right)
\ln (k\eta_1)\right].
\end{eqnarray}
This holds for $\frac{q}{k}<a^{-1}$.

\subsubsection*{Case (i) matched to (iv):}
\begin{eqnarray}
c_{-}e^{i\frac{q\eta_1}{2}}\sim
-\frac{\sqrt{2}}{\pi\sqrt{q\eta_1}}
\left[1+(1+iq\eta_1)\ln (k\eta_1)\right],
\end{eqnarray}
which holds for
$
a^{-1}<\frac{q}{k}<\left(\frac{k_1}{k}\right)^{-\beta}.
$

\subsubsection*{Case (ii) matched to (iv):}
\begin{eqnarray}
c_{-}e^{i\frac{q}{2}\eta_1}\sim
-\frac{\sqrt{2}}{\pi\sqrt{q\eta_1}}
\left[\beta+1+(1+iq\eta_1)\ln (\kappa q\eta_1)\right].
\end{eqnarray}
This holds for
$\frac{q}{k}>\left(\frac{k_1}{k}\right)^{-\beta}$.

\subsubsection*{Case (ii) matched to (iii):}
Clearly, this possibility is not allowed within our set-up, where the
extra (internal) dimensions are collapsing while the external four
dimensions are expanding. This implies $\beta < 0$, and therefore case
(ii) requires $q/k > 1$, while case (iii) holds for the opposite
limit, namely $q/k < 1$. Therefore, our scenario does not allow the
matching of cases (ii) and (iii).

\vspace{1cm}

In order to compare with observations it is useful to calculate the
spectral energy density.
The total energy density of created particles is given by
\begin{eqnarray}
\rho=2\frac{R^n}{(2\pi)^{n+3}}\int\left[\frac{k^2}{a^2}+\frac{q^2}{b^2}\right]^{\frac{1}{2}}
|c_{-}|^2{\rm d}V_{\rm phys} 
\end{eqnarray}
and ${\rm d}V_{\rm phys}={\rm d}V_{\rm com}/(a^3b^n)$,
where it is assumed that the comoving volume consists of
two spheres, i.e.
$${\rm d}V_{\rm com}=\frac{2\pi^{\frac{d}{2}}}{\Gamma(\frac{d}{2})}
k^{d-1}{\rm d}k\wedge\frac{2\pi^{\frac{n}{2}}}{\Gamma(\frac{n}{2})}
q^{n-1}{\rm d}q~,$$ which for $d=3$ implies
\begin{eqnarray}
\rho=\frac{R^n}{(2\pi)^{n+3}}16\frac{\pi^{1+\frac{n}{2}}}{\Gamma(\frac{n}{2})}
a^{-4}b^{-n}
\int\left[1+\left(\frac{a}{b}\right)^2
\left(\frac{q}{k}\right)^2\right]^{\frac{1}{2}}
k^3q^{n-1}|c_{-}|^2{\rm d}k {\rm d}q.
\end{eqnarray}
Hence the spectral energy density $\rho(k)=d\rho/d\log k$ is given by
\begin{eqnarray}
\rho(k)=\frac{R^n}{(2\pi)^{n+3}}16\frac{\pi^{1+\frac{n}{2}}}{\Gamma(\frac{n}{2})}
\left(\frac{k}{a}\right)^4
\left(\frac{k}{b}\right)^{n}
\int {\rm d}Y Y^{n-1}\left[1+\left(\frac{a}{b}\right)^2
Y^2\right]^{\frac{1}{2}}|c_{-}|^2~,
\end{eqnarray}
where $Y\equiv q/k$.

As explained above there are three contributions to the
spectral energy density. The notation $\rho_{(i)-(iii)}$
stands for the contribution resulting from the 
Bogoliubov coefficients calculated from the matching
of cases (i) and (iii), and in a similar way for the other cases. 
For $\rho_{(i)-(iii)}$,
the integral 
$$\int_0^{1/a}{\rm d}YY^{n-1}\left[1+\left(\frac{a}{b}\right)^2Y^2
\right]^{1/2}|c_{-}|^2$$ has to be calculated. For 
$\rho_{(i)-(iv)}$, the corresponding one. In order to calculate
$\rho_{(ii)-(iv)}$, we introduce an upper cut-off in the 
integral, i.e. 
$$\int_{(k_1/k)^{-\beta}}^{q_{\rm max}/k}
{\rm d}YY^{n-1}\left[1+\left(\frac{a}{b}\right)^2Y^2
\right]^{1/2}|c_{-}|^2~.$$
At the transition time $\eta=-\eta_1$ the metric is continuous but
its first derivative is not. This leads to the so-called
sudden transition approximation. 
It basically means that for the modes with periods
much greater than the duration of the transition phase, the
transition can be considered as instantaneous.
It is known that this type of approximation leads
to an ultra-violet divergency \cite{GV}\cite{hu}. 
This justifies the introduction of an upper cut-off $q_{\rm max}$.
Neglecting sub-leading terms for the total spectral energy density, 
$\rho(k)=\rho_{(i)-(iii)}+\rho_{(i)-(iv)}+\rho_{(ii)-(iv)}$, 
the following expression is found 
\begin{eqnarray}
\rho(k) &\sim &
R^n\frac{2^{-n}}{a^{4+n}}\frac{\pi^{-4-\frac{n}{2}}}
{n\Gamma(\frac{n}{2})}k_1^{4+n}
\left(\frac{k}{k_1}\right)^{3+n}
\left(\ln\frac{k}{k_1}\right)^2\nonumber\\
&+&R^n\frac{2^{2-n}}{a^3}\frac{\pi^{-4-\frac{n}{2}}}
{n\Gamma(\frac{n}{2})}k_1^{4+n}
\left(\frac{k}{k_1}\right)^{3+n(\beta+1)}
\left(\ln\frac{k}{k_1}\right)^2\nonumber\\
&+&R^n\frac{2^{2-n}}{a^3}\frac{\pi^{-4-\frac{n}{2}}}
{n\Gamma(\frac{n}{2})}k_1^{4+n}
\left(\frac{k}{k_1}\right)^{3}
\left(\frac{q_{max}}{k_1}\right)^n
\left(\ln\kappa\frac{q_{max}}{k_1}\right)^2.
\end{eqnarray}
This can be expressed in terms of the fractional energy density
(dimensionless parameter),
$\Omega(k)\equiv \rho(k)/\rho_{\rm c}$, where the 
critical energy density $\rho_{\rm c}$ is given in terms of the
$D$-dimensional Planck mass, $M_{\rm D}$,  as
 $\rho_{\rm c}=[3/(8\pi)] R^nM_{\rm D}^{n+2} H^2$.
We thus find that the fractional energy density of produced gravitons
$\Omega_{\rm GW} (\eta)$ at a given time $\eta$,
with respect to the fraction of critical energy density in radiation,
$\Omega_{\gamma}(\eta)$, defined as
$\Omega_{\gamma}(\eta)=(H_1/H)^2(a_1/a)^4$, is
\begin{eqnarray}
\Omega_{\rm GW} (\eta)&\sim&{\cal A}_1a^{-n}\left(\frac{H_1}{M_{D}}\right)^{2+n}
\Omega_{\gamma}(\eta)\left(\frac{k}{k_1}\right)^{3+n}
\left(\ln\frac{k}{k_1}\right)^2\nonumber\\
&+&{\cal A}_2a\left(\frac{H_1}{M_{D}}\right)^{2+n}
\Omega_{\gamma}(\eta)\left(\frac{k}{k_1}\right)^{3+n(\beta+1)}
\left(\ln\frac{k}{k_1}\right)^2\nonumber\\
&+&{\cal A}_2a\left(\frac{H_1}{M_{D}}\right)^{2+n}
\Omega_{\gamma}(\eta)\left(\frac{k}{k_1}\right)^{3}
\left(\frac{q_{max}}{k_1}\right)^n\left(\ln\kappa\frac{q_{max}}{k_1}
\right)^2.
\label{om}
\end{eqnarray}
The constants ${\cal A}_i$ contain all the numerical factors.
Furthermore, it was assumed that the maximal frequency $k_1$ 
in the four-dimensional space-time is of the same order as 
the curvature scale at the time of 
transition, i.e. $k_1\sim H_1$.

This expression can be compared with the one for no excited internal
momenta.  It can be easily recovered from the matching of case (i) to
(iii), without carrying out the integration.  It is found that
\begin{eqnarray}
\Omega_{{\rm GW}\;|\;q=0} (\eta)={\cal A} \left(\frac{H_1}{M_{4}}\right)^{2}
\Omega_{\gamma}(\eta)\left(\frac{k}{k_1}\right)^{3}
\left(\ln\frac{k}{k_1}\right)^2.
\end{eqnarray}

As it can be seen from the expression for the 
fractional energy density, Eq.~(\ref{om}),
the effect of the momenta in the extra dimensions is two-fold.
On the one hand, there are two contributions which are growing 
with the scale factor,
and therefore might soon dominate the energy density of the universe
and overclose it. On the other hand, the cut-off $q_{\rm max}$ has to be 
chosen in such a way that its contribution does not dominate, i.e.
$q_{\rm max}\sim (\beta+1)k_1$. Thus, the cut-off of the momenta
in the extra space has to be less than the one in the
four-dimensional space-time, $q_{\rm max}<k_1$.

There are several constraints on the gravitational wave spectrum.  The
strongest ones come from big bang nucleosynthesis and the prevention
of recollapse of the universe.  Both of these hold for the whole range
of frequencies.  Big bang nucleosynthesis successfully predicts the
production of light elements. However, everything depends crucially on
the expansion rate at the time of nucleosynthesis. This in turn is
determined by the total energy density at that time.  Therefore, any
extra contributions at that time are constrained.  The total energy
density of gravitational waves at the present time has to satisfy
\cite{mm}\cite{bbn}
\begin{eqnarray}
\int_{f=0}^{f=\infty}{\rm d}(\log f)
~h_0^2~\Omega_{\rm GW}(f)~\leq~0.227~\Omega_{\gamma}~,
\end{eqnarray}
where $0.5<h_0<0.85$ parametrizes the experimental uncertainty for 
the present value of the Hubble parameter $H_0$.
One might argue that, as a first approximation,
$h_0^2~\Omega_{\rm GW}(f)\leq~0.227~\Omega_{\gamma}$, which leads to
\begin{eqnarray}
{\cal A}_2 
a_0\left(\frac{H_1}{M_{D}}\right)^{2+n} < 0.227~.
\label{req}
\end{eqnarray}
Furthermore, $a_0$ is given by
$$
a_0=\Omega_{\gamma,0}^{-1/4}\left(H_1\over H_0\right)^{1/2} \sim
5\times10^{31}\left({H_1\over M_{\rm 4}}\right)^{1/2}~,
$$
where the subscript 0 denotes the present epoch.  Assuming that the
standard behaviour of radiation holds up to the transition scale, the
Hubble parameter $H$ and the temperature $T$ are related by (see for
example \cite{KT})
\begin{eqnarray}
H=1.66g_{*}^{\frac{1}{2}}\frac{T^2}{M_4}~,
\label{HT}
\end{eqnarray}
where $g_{*}(T)$ counts the total number of effectively massless
degrees of freedom. For temperatures $T>$300 GeV, one has
$g_{*}=106.75$. Using Eq.~(\ref{HT}) for the value of the Hubble parameter
at the time of transition, $H_1$, the bound given by Eq.~(\ref{req})
can be written in terms of the temperature at the transition, i.e. at
the beginning of the radiation dominated FLRW stage.  The maximal
temperature $T_1$ is given when the bound is satisfied. It is found to
be
\begin{eqnarray}
T_1=\left[3I\frac{(1.211)^{3+n}}{5}
\frac{10^{32+22n}}
{[1.66g_{*}^{1/2}]^{\frac{5}{2}+n}2^{5-n}}
 \pi^{3+\frac{n}{2}}
      n\Gamma(\frac{n}{2})
\right]^{\frac{1}{5+2n}}
\left(\frac{M_D}{\rm 1 TeV}
\right)^{\frac{2+n}{5+2n}}{\rm GeV}~,
\label{T_RH}
\end{eqnarray}
where $I=0.227$ is our bound. This is shown in Figure 2.
\begin{figure}
\centerline{\epsfxsize=2.5in\epsfbox{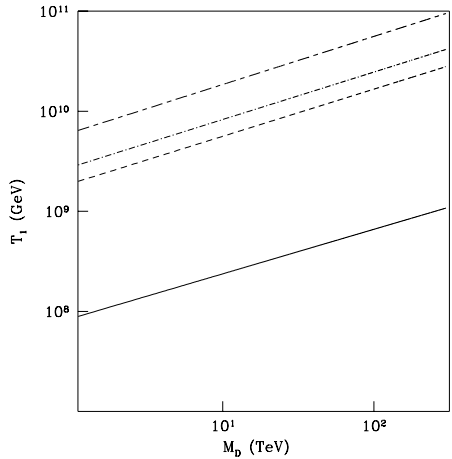}}
\caption{The maximal temperature 
at the beginning of the radiation dominated FLRW stage as a function
of the $D$-dimensional Planck mass. The solid line shows $n=2$
extra dimensions, the long dashed one is for $n=6$, the dashed-dotted
line for $n=7$ and the long/short-dashed one for $n=10$. 
}
\label{fig2}
\end{figure}
It should be noted that the range of the $D$-dimensional Planck mass 
is constrained by the upper size of the extra dimensions, 
$R_{\rm max}\sim 1{\rm mm}$. This gives the constraint 
\begin{eqnarray}
M_D\geq\left(2\times 10^{(32/n)-16}\right)^{1/(1+\frac{2}{n})} {\rm TeV}~.
\label{M_D}
\end{eqnarray}
Therefore, there is a lower bound on $T_1$.
The minimal value of $M_{\rm D}$ and the minimal
value of $T_1$ as a function of the number $n$ of
extra dimensions are shown in Figure 3.
\begin{figure}
\centerline{\epsfxsize=2.5in\epsfbox{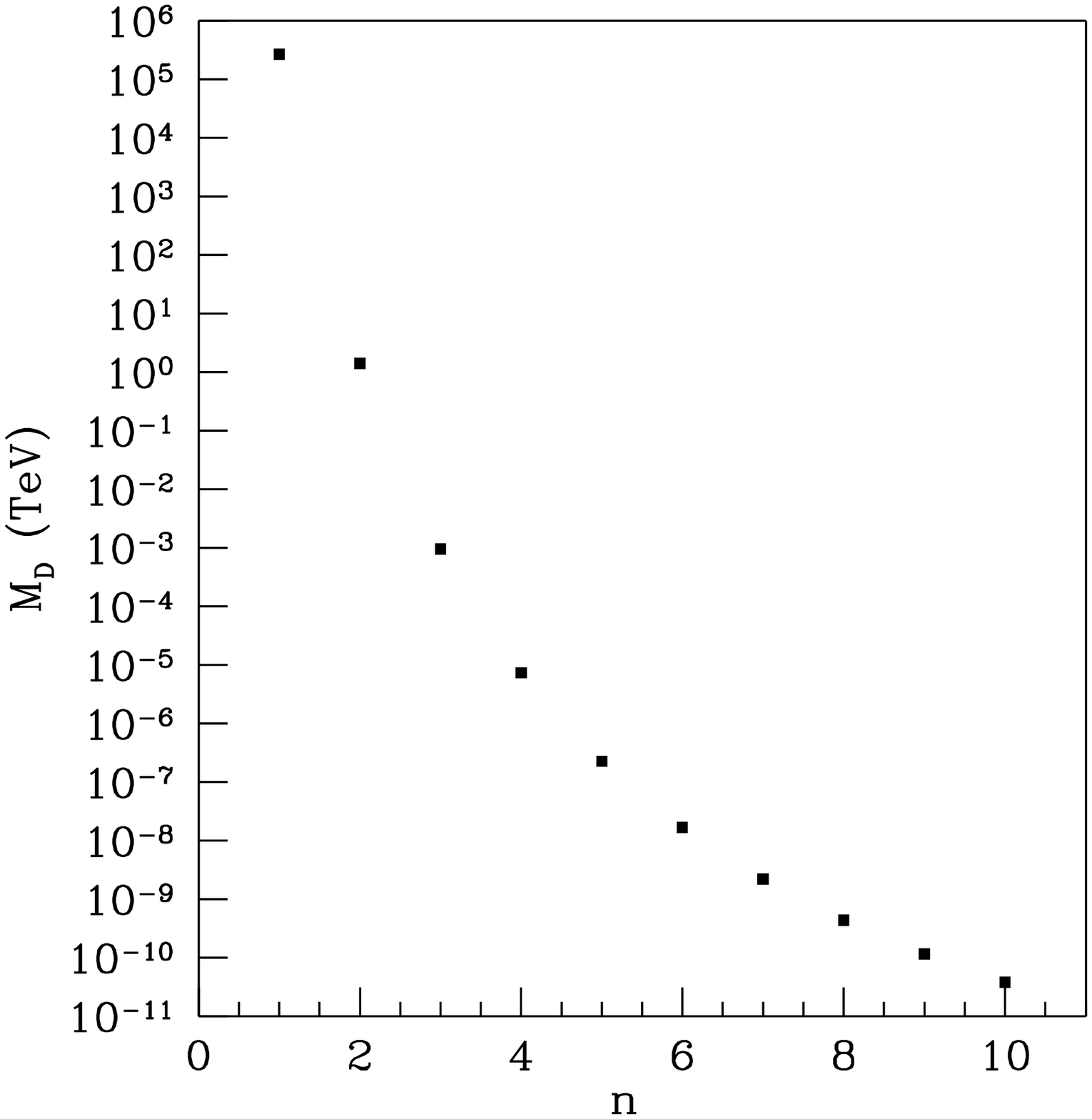} \hspace{1cm}
              \epsfxsize=2.5in\epsfbox{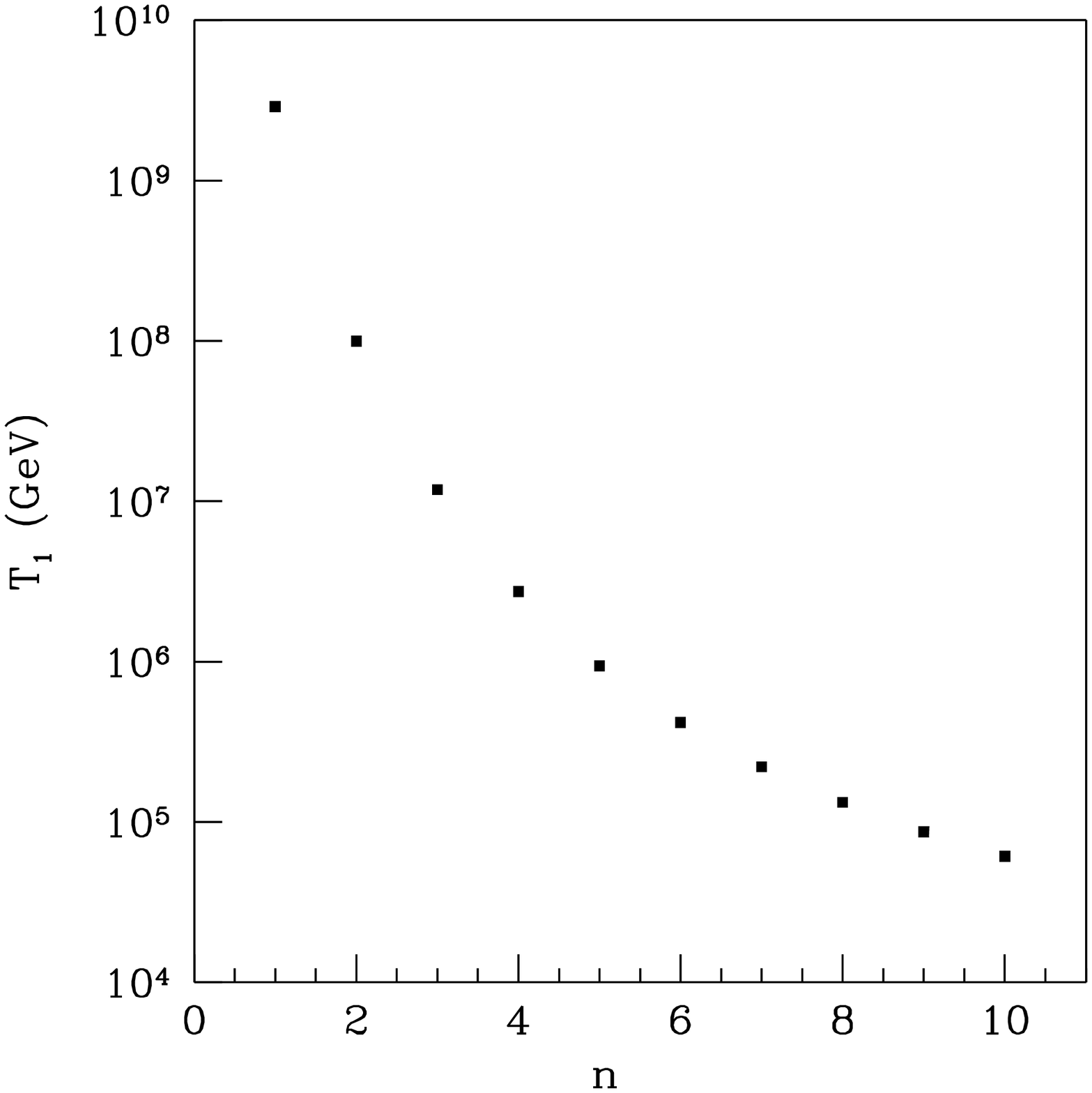}}
\caption{In the left panel the allowed
minimal value of the $D$-dimensional Planck mass $M_{\rm D}$
is shown as a function of the number of extra dimensions
$n$. In the right panel the minimal value of the temperature
at the beginning of the radiation dominated stage, $T_1$,
is shown as a function of $n$. 
}
\label{fig3}
\end{figure}
Assuming that $T_1$ is the same as the reheating temperature $T_{\rm
RH}$, Eqs.~(\ref{T_RH}) and (\ref{M_D}) imply a lower bound on $T_{\rm
RH}$.  More precisely, for one extra dimension $(n=1)$, the lower
bound for the $D$-dimensional Planck mass is $M_{\rm D}>10^5 {\rm TeV}$,
leading to a lower bound for the reheating temperature $T_{\rm RH}>
3\times 10^9 {\rm GeV}$.  
In a similar way, for two extra dimensions
$(n=2)$, one obtains $M_{\rm D}>1 {\rm TeV}$ and $T_{\rm RH}> 10^7 {\rm
GeV}$. Finally, for $n=3$ one has $M_{\rm D}>10^{-3} {\rm TeV}$ and
$T_{\rm RH}> 5\times 10^6 {\rm GeV}$. 

Big-bang nucleosynthesis must proceed in the standard way, implying
that the reheating temperature must be higher than $\sim 10 {\rm
MeV}$.  This requirement is satisfied in our model, as the above
analysis shows.  Standard inflationary models, on the other hand, lead
to a higher reheating temperature than the lower bounds we found
above. However, one must keep in mind that above we only gave the
lower bounds, while the actual reheating temperature can be much
higher indeed. In addition, a higher value can be easily achieved
increasing the value of the $D$-dimensional Planck mass, provided it
always satisfies the constraint given in Eq.~(\ref{M_D}). Finally,
reheating from a higher number of collapsing extra dimensions favours
larger values of $M_D$, and thus smaller extra dimensions.

The other strong constraint comes from the requirement that the
additional energy density should not overclose the universe. Thus, the 
condition $\Omega_{\rm GW}(\omega)<1$ has to hold for all frequencies. 
However, if the big bang nucleosynthesis constraint is
satisfied, then indeed $\Omega_{\rm GW}<1$.

There are additional constraints coming from large scales as probed by
the cosmic microwave background explorer (COBE).  Strong perturbations
can change the gravitational potential experienced by the microwave
background photons. This causes a red-shift of these photons and hence
a fluctuation in temperature of the cosmic microwave background
\cite{mm}.  There is, in addition, a bound due to pulsars whose signal
is delayed, if a gravitational wave passes through between the earth
and the pulsar \cite{mm}. However, both these constraints apply
only for a range of frequencies and they are in general less tight
than the ones mentioned earlier.  Furthermore, since the spectrum is
increasing with increasing frequency, the most stringent bounds will
come from large frequencies, i.e. small wavelengths.

In conclusion, we find that the excited internal modes are compatible
with observations, provided the temperature at the transition time to
an effectively four-dimensional universe, is less than its maximum,
given in Eq.~(\ref{T_RH}).  Thus, recalling that $H_1 \sim T_{\rm
RH}^2/M_4$, $H_1\sim k_1$, $q_{\rm max}<k_1$ and $q\sim R^{-1}$, this
implies that the size $R$ of the extra dimensions has to be $R >
M_4/T_{\rm RH}^2$.  As we have shown, the lower bound of the reheating
temperature depends on the number $n$ of extra dimensions and the
value of the $D$-dimensional Planck mass $M_{\rm D}$.  With values of
$M_{\rm D}$, such that the constraint Eq.~(\ref{M_D}) is satisfied,
one can easily check that our analysis leads to the requirement that
the size of the internal space must be much larger than the
four-dimensional Planck length
\begin{equation}
R\gg \ell_{{\rm Pl}}~,
\end{equation}
where $\ell_{{\rm Pl}}$ is the four-dimensional 
Planck length $(\ell_{{\rm Pl}} \sim 10^{-19} 
{\rm GeV}^{-1})$.

Our model could very well be realized within the context of the
pre-big-bang scenario \cite{pbb}, a particular model of inflation
inspired by the duality properties of string theory. In the
pre-big-bang scenario, perturbations of Kalb-Ramond axions can provide
a quasi-scale-invariant (Harrison-Zel'dovich) spectrum of large
angular scale cosmic microwave background (CMB) anisotropies
\cite{dgsv}.  An extension of this mechanism (for massless axions) to
the region of the acoustic peaks, showed that a consistency with the
current CMB data requires that the internal dimensions contract at a
rate faster than the rate at which the external dimensions expand
\cite{mvdv}.  Therefore, combining our model with the one where the
universal axion of string theory triggered the CMB anisotropies, we
get further constraints on the extra dimensions of string theory.
However, this will be left for future work.

\section{Conclusions}
We have discussed graviton production in the context of a higher
dimensional model, which from a multi-dimensional phase where the
extra dimensions are contracting and the external dimensions are
expanding, enters into an effectively four-dimensional universe with
static extra dimensions.  The momenta in the extra space were taken
into account.  To find an estimate of their contribution to the energy
density in the four-dimensional space-time, the momenta in the internal 
space were integrated out.

The contributions of the internal momenta tend to dominate the
gravitational wave spectrum, up to the point of forcing the universe
to recollapse at an early stage.  However, one can assume that there
is an upper cut-off for the internal momenta, given by the maximal
frequency of the four-dimensional (external) space-time.  The
gravitational wave spectrum is constrained by observations.  With the
assumption that the standard relation between the Hubble parameter and
the temperature holds up to the transition time, we derived an upper
bound on the temperature at the beginning of the radiation dominated
era.  Using that Newtonian gravity has been tested sucessfully down to
scales of the order of 1 mm, one obtains a lower bound on the
temperature at the transition scale, which depends on the number of
extra dimensions.  This lower bound on the reheating temperature is,
in general, lower than the reheating temperature of standard
inflationary models. However, it can be raised if one assumes smaller
extra dimensions, which lead to a higher $D$-dimensional Planck mass.
Our final conclusion is that the size $R$ of the internal space must
be much larger than Planck length $\ell_{\rm Pl}$, namely $R\gg
\ell_{\rm Pl}$.

\section{Acknowledgements}
It is a pleasure to thank R.~Durrer for enlightening discussions and
collaboration during the early stages of this work.  We are also
grateful to R.~Brandenberger for careful reading of the manuscript and
for various illuminating comments. We also thank M.~A.~V\'azquez-Mozo
for discussions. K. E. K. thanks the department of theoretical physics at
Geneva University where the major part of this work was done.
While at Geneva University K.~E.~K was supported by the Swiss NSF. 
M.~S. acknowledges
financial support from E.L.K.E. ({\sl Special Account for Research}), University
of Athens, Hellas.

\end{document}